\documentclass[pra,twocolumn,aps,showpacs]{revtex4-1}
\usepackage{graphicx}
\usepackage{color}

\begin{document}
\title{Transverse localization in nonlinear photonic lattices with second-order coupling}

\author{M. Golshani}
\affiliation{Department of Physics, Sharif University of Technology, Tehran, Iran}

\author{A. R. Bahrampour}
\affiliation{Department of Physics, Sharif University of Technology, Tehran, Iran}

\author{A. Langari}
\affiliation{Department of Physics, Sharif University of Technology, Tehran, Iran}

\author{A. Szameit}
\affiliation{Institute of Applied Physics,
Abbe Center of Photonics, Friedrich-Schiller-University Jena, Max-Wien-Platz 1, 07743 Jena, Germany}


\begin{abstract}
We investigate numerically the effect of long-range interaction on the transverse localization of light. To this end, nonlinear zigzag optical
waveguide lattices are applied, which allows precise tuning of the second-order coupling. We find that localization is hindered by coupling between
next-nearest lattice sites. Additionally, (focusing) nonlinearity facilitates localization with increasing disorder, as long as the nonlinearity is
sufficiently weak. However, for strong nonlinearities, increasing disorder results in \textit{weaker} localization. The threshold nonlinearity, above
which this anomalous result is observed grows with increasing second-order coupling.
\end{abstract}

\maketitle

\section{Introduction}

The localization of waves in disordered lattice systems, as proposed in 1958 by P.W. Anderson \cite{Anderson1958}, is a fascinating feature of energy
transport. It was predicted that, when disorder is introduced to a periodic system, the extended Bloch eigenmodes may convert to exponentially
localized states, leading to metal-insulator transition \cite{MITransitionBook}. The interference between multiple scattered electronic waves is the
origin of Anderson localization. Although, the Anderson localization was initially introduced in solid-state physics, due to its wave nature based on
interference effects, this concept could be applied to other waves such as light \cite{LocalizationLight,SeeLightin3D,TL2D}, matter
\cite{localizationMatterWaves}, and even sound \cite{LocalizationSound}. Notably, all of these studies do not consider the impact of non-nearest
lattices sites on localization. However, in some systems like biomolecules \cite{BioMol} and polymer chains \cite{Polymer}, long-range interactions
in the lattice become important and cannot be neglected.

In recent years, coupled optical waveguide lattices provide an excellent platform for study of transverse localization of light \cite{TL,TL2D,TL1D}.
In particular, optical lattices fabricated using the femtosecond (FS) laser writing technology are particularly useful to study various effects
associated with disorder \cite{TL1DOffdiagonal,offdiagonalmartin,Stuetzer,NaetherDisorderedBoundary,NaetherTransition1D2D}. Besides these
experimental studies of light localization in disordered photonic lattices, there is a large number of numerical or theoretical literature on the
different aspects of these systems \cite{TheoryandNumericalTL}. For instance, the competition between disorder and nonlinearity \cite{TLNonlinear},
dependence of localization on input beam profile \cite{TLInputProfile}, transverse localization with dimensionality crossover \cite{TLDCross}, and
the effect of the excited site number on the wave-packet localization \cite{MolinaPRE2012} have been studied previously.

In our work, we study numerically the impact of interactions between non-nearest lattice sites on Anderson localization in a disordered system in the
linear and nonlinear regime. To this end we employ zigzag lattices of optical waveguides
\cite{NikosPRE2002,ZigzagArray,SzameitSOC2008,DiscreteSolitons}, which are particularly useful to study second-order coupling (SOC) between the
lattice sites.

This paper is organized in four sections. Section II contains the theoretical considerations of disordered zigzag arrays of optical waveguides, and
the definition of standard quantities for investigation of transverse localization in these systems. In section III, the results of numerical
simulations and discussions will be presented, and finally, conclusions will be presented in section IV.

\section{Theoretical Considerations}

Our system consists of a zigzag array of identical single-mode circular optical waveguides (Fig. 1). We assume that the angles
between adjacent arms are the same ($\theta$), while the distance between two successive waveguides are selected randomly from
an uncorrelated exponential probability distribution. According to the exponential dependence of coupling coefficient with
waveguide separation \cite{SzameitControlofEvanescentCoupling2007}, this will lead to a uniform random distribution of coupling
constants between adjacent guides.

In the coupled mode approximation, the evolution equation for the complex amplitude of electric field at the jth waveguide $E_{j}$ is
\cite{NikosPRE2002}
\begin{eqnarray}
  \label{eq1}
  -i\frac{d E_j}{dz}&=&t_1^{+}(j) E_{j+1}+t_1^{-}(j) E_{j-1}
        +t_2^{+}(j) E_{j+2}
    \nonumber \\
        & &+t_2^{-}(j) E_{j-2}+\chi |E_j|^2 E_j
\end{eqnarray}
with $j=1,\ldots,N$, where $t_1^{\pm}(j)$ are the first-order coupling (FOC) coefficients between guides $j$ and $j\pm1$, $t_2^{\pm}(j)$ are the SOC
coefficients between non-nearest neighbors (NNN) $j$ and $j\pm2$, and $\chi$ is the nonlinear Kerr constant (we consider only
focusing nonlinearities $\chi\geq 0$).

In lossless systems $t_m^{+}(j)=t_m^{-}(j+m), m=1,2$ \cite{SzameitControlofEvanescentCoupling2007}. The FOC coefficients $t_1^{\pm}(j)$ are random
numbers with uniform distribution in the interval $[t_0(1-\Delta),t_0(1+\Delta)]$, where $0\leq \Delta<1$ is the disorder strength. We set the
average distance between successive waveguides ($d_w$) to 18 $\mu m$, corresponding to $t_0\simeq 0.15 \ mm^{-1}$ (for $\lambda =780 \  nm$)
\cite{SzameitControlofEvanescentCoupling2007}. For each realization $\{d_j \}$, the SOC coefficients $t_2^{\pm}(j)$ can be determined by the NNN
separation \cite{SzameitControlofEvanescentCoupling2007}
\begin{equation}
\label{eq2}
   d_{j,j+2}=\sqrt{d_j^2+d_{j+1}^2-2d_j d_{j+1}\cos\theta}
\end{equation}
The relative coupling strength $\alpha$, which is defined as $t_2^{\pm}/t_1^{\pm}$ at $\Delta=0$, can be controlled with the angle
$\theta$ in the Fig. 1. Not that at $\theta <60^{\circ}$ one finds $\alpha >1$. The nonlinearity in the system can be tuned by the
Kerr parameter $\chi$, and $\chi=0$ corresponds to purely linear dynamics.

To investigate the effect of nonlinearity and SOC on \textcolor{black}{Anderson} localization, we solve Eqs. (\ref{eq1}) numerically by the
Runge-Kutta-Fehlberg method \cite{RKFMethodBook}, with single site excitation ($E_j(z=0)=\delta_{j,j_0}$) as initial condition. To consider finite
size system effects, we use fixed boundary conditions, i.e., $E_j=0$ for $j=-1,0,N+1 $ and $N+2$. Under this condition the norm $\sum_{j=1}^N
|E_j|^2$ is conserved. To peruse transverse localization, we use two measures: the
dimensionless
transverse localization length (TLL) $\ell$, which is obtained by
fitting an exponential function $|E_j|^2=|E|_{max}^2 e^{-|j-j_{c}|/\ell}$ on the localized profile, and the the effective width $w_{eff}$
\cite{TL2D}:
$$ w_{eff}=\sqrt{\frac{(\sum_{j=1}^N |E_j|^2)^2}{\sum_{j=1}^N |E_j|^4}}$$ that illustrates the formation of a localized state during the propagation.
Due to the statistical nature of Anderson localization, both quantities are averaged over 5000 different realizations.

\begin{figure}[t!]
\includegraphics[angle=0, width=0.9\linewidth]{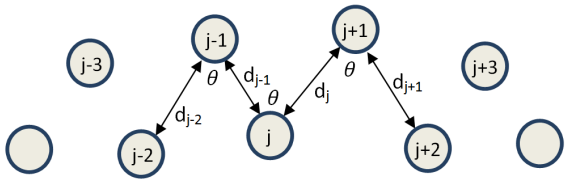}
\caption{Zigzag array of optical waveguides. The distance between guides j and j+1 is denoted by $d_{j}$, and the angle between adjacent arms is $\theta$.}
\end{figure}

\begin{figure}[t!]
\includegraphics[angle=0,height=1.05\linewidth, width=0.9\linewidth]{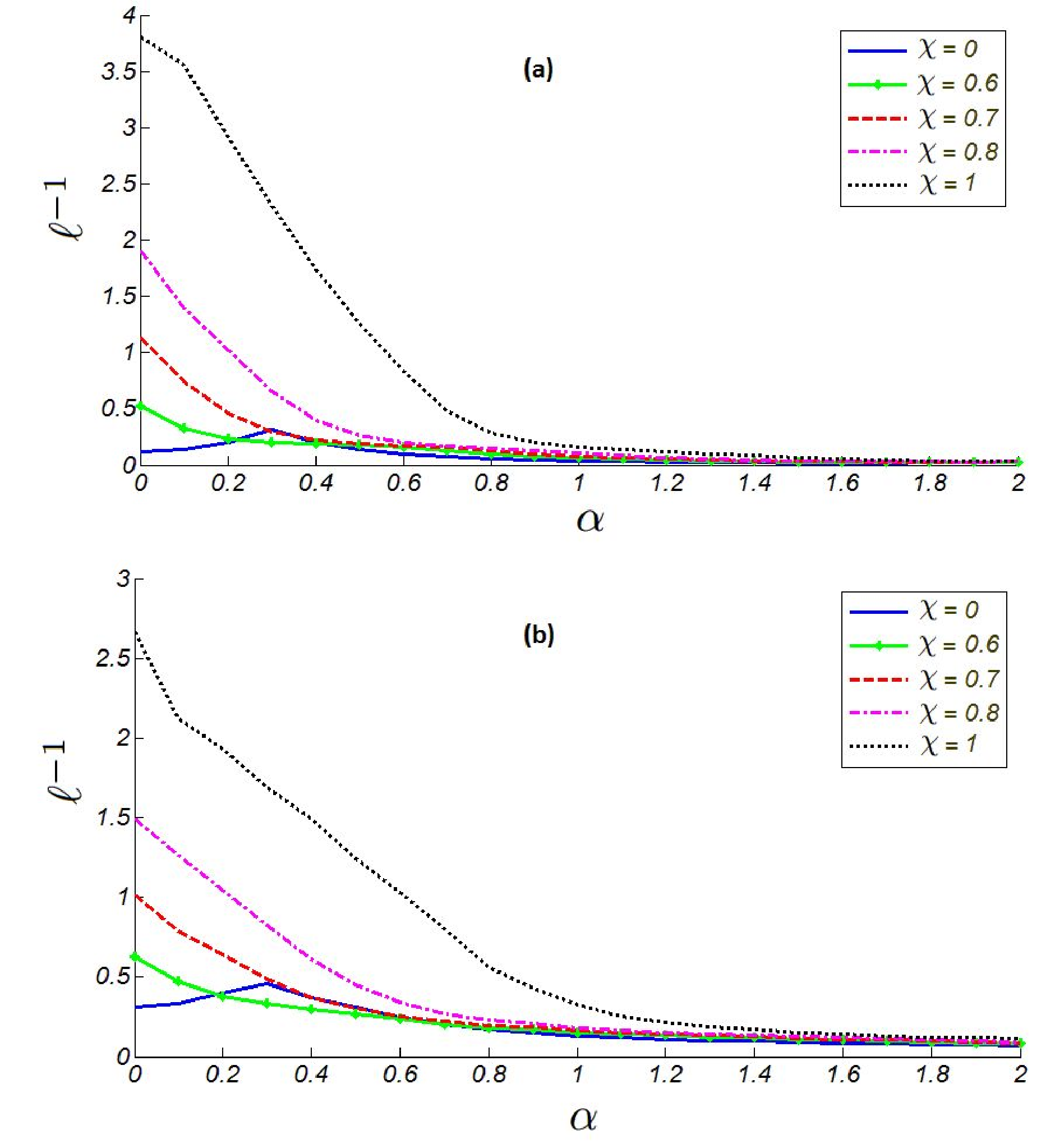}
\caption{(color online) Inverse of transverse localization length $\ell^{-1}$ versus relative coupling strength $\alpha$ for different values of the
nonlinear parameter $\chi$ ($mm^{-1}$). The excitation is at site $j_0=50$. (a) $\Delta=0.4$, (b) $\Delta=0.7$.}
\end{figure}
\begin{figure}[t!]
\includegraphics[angle=0,height=0.6\linewidth, width=0.9\linewidth]{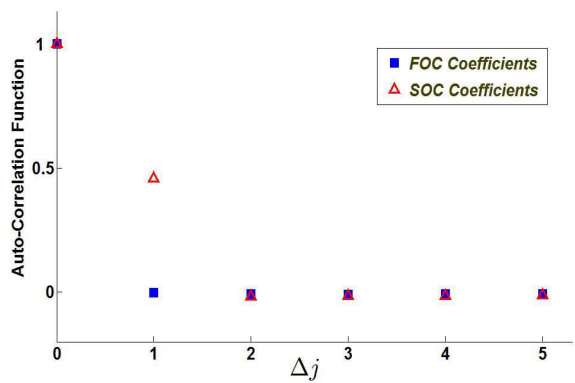}
\caption{(color online) Autocorrelation function of FOC and SOC coefficients for $\Delta=0.7$.}
\end{figure}

\section{Results and Discussions}

For our simulation, we consider zigzag arrays containing $N=100$ waveguides each has $100 \ mm$ length. \textcolor{black}{In Fig. 2, we show the
inverse TLL ($\ell^{-1}$) at the end of the waveguide array as a function of $\alpha$ for two disorder strengths $\Delta = 0.4,0.7$ and several
nonlinear parameters, when the lattice bulk was excited ($j_0=50$). The main result of our simulations is that, in general, SOC impedes localization
and increases the transverse spreading of the wave function.} According to the band random matrix theory \citep{RandomBandMatrixs},
it is well known that the localization length of the eigen states, at large values of the band width, increases with the square of the band width.
The increase of the transverse spreading can be explained in terms of a modified band structure of the Hamiltonian matrix, in the presence of SOC. In
the presence of coupling to the second-nearest neighbor, the Hamiltonian matrix of the system will be changed from tridiagonal form to a
pentadiagonal one, hence the direct coupling paths for spreading of the wave packet will be increased. Moreover, short-range correlations between the
lattice sites, introduced by SOC, is another important feature of this system which may affect on the system localization \citep{Correlation2001}. In
Fig. 3, we plot the autocorrelation functions of the FOC and SOC at $\Delta = 0.7$, which indicates that although the FOC coefficients are chosen
from a uniform uncorrelated random sequence, the SOC coefficients have positive short-range correlation. This is due to the presence of the joint arm
$d_{j+1}$ in Eq. \ref{eq2} for the expression of $t_2^{+}(j)$ and $t_2^{+}(j+1)$. We find that the autocorrelation functions do not show any
noticeable change with the variation of $\alpha$. However, by decreasing the disorder strength $\Delta$ the value at $\Delta j = 1$ increases
slightly.

To investigate the effect of correlation on the localization length, we repeat our simulations but with uncorrelated random hopping
terms. It is important to notice that these uncorrelated random terms are chosen from the same probability distributions related to the model under
study. Figs. 4 and 5 show the inverse TLL ($\ell^{-1}$) at the end of waveguide array as a function of $\alpha$ for two cases of correlated and
uncorrelated SOC coefficients, for two disorder strengths $\Delta=0.4, 0.9$ and nonlinear parameters $\chi=0.1\; mm^{-1}$ and $\chi=1\; mm^{-1}$. We
find that site correlation effects appear only at high disorder strengths for small nonlinear parameters. In addition, in these regimes, correlation
will decrease TLL ($\ell$) slightly \cite{Correlation2009,Correlation2012}.
\begin{figure}[t!]
\includegraphics[angle=0,height=1\linewidth, width=0.9\linewidth]{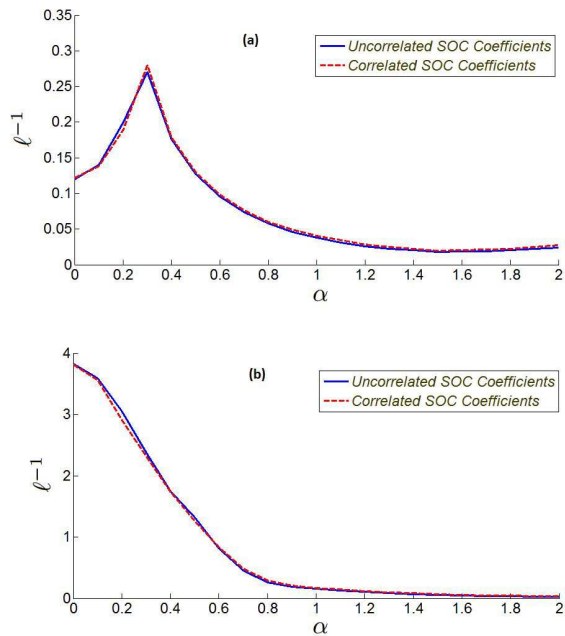}
\caption{(color online) Inverse of transverse localization length $\ell^{-1}$ versus relative coupling strength $\alpha$ for
two case of correlated and uncorrelated SOC coefficients in the weak disorder regime $\Delta=0.4$. Excited site number
is $j_0=50$, and (a) $\chi=0.1\ mm^{-1}$, (b) $\chi=1\ mm^{-1}$.}
\end{figure}
\begin{figure}[t!]
\includegraphics[angle=0,height=1\linewidth, width=0.9\linewidth]{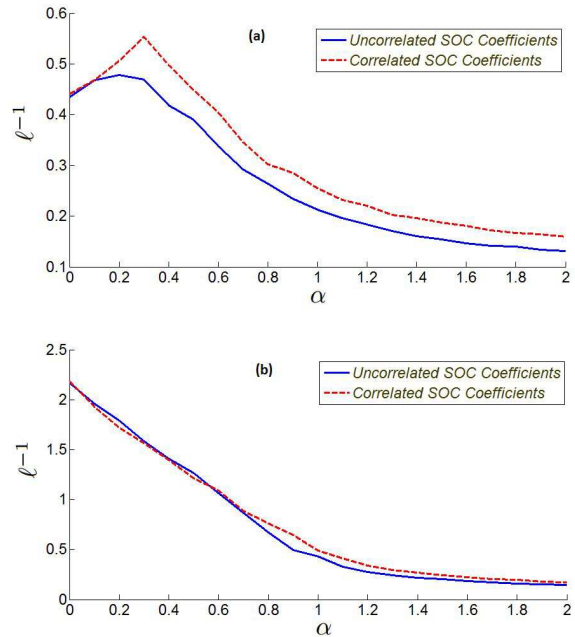}
\caption{(color online) Inverse of transverse localization length $\ell^{-1}$ versus relative coupling strength $\alpha$
for two case of correlated and uncorrelated SOC coefficients in the strong disorder regime $\Delta=0.9$. Excited site
number is $j_0=50$, and (a) $\chi=0.1\ mm^{-1}$, (b) $\chi=1\ mm^{-1}$.}
\end{figure}

Two selected propagation images for small ($\alpha=0.3$) and high ($\alpha=2$) SOC strengths, shown in Fig. 6, illustrate our
findings. The plots explicitly demonstrate the enhanced expansion of the propagating wave packet for higher $\alpha$,
i.e., with increasing SOC.

\begin{figure}[t!]
\includegraphics[angle=0,height=0.55\linewidth, width=0.9\linewidth]{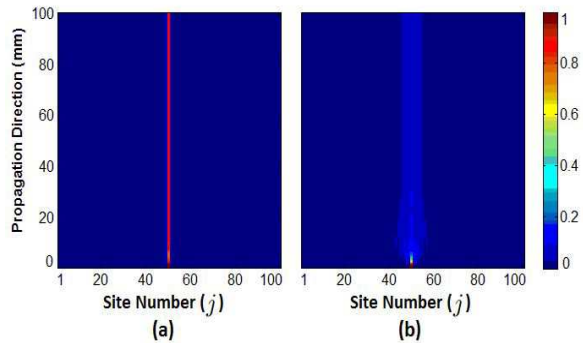}
\caption{(color online) Propagation images for $\Delta=0.7$,  $\chi=1\ mm^{-1}$, $j_0=50$ and (a) $\alpha=0.3$, (b) $\alpha=2$. }
\end{figure}

Another important result of our simulations is the enhancement of localization by increasing nonlinearity (see Fig. 7). In fact,
one identifies two distinct nonlinear regimes, where the inverse TLL $\ell^{-1}$ has small and large values \cite{MolinaPRE2012}, respectively. When
the nonlinear parameter $\chi$ is below some characteristic value $\chi_c$ (see Fig. 7) (weakly nonlinear regime), localization growth slightly with
increasing nonlinearity. In contrast, when the nonlinear parameters is larger than the critical value $\chi>\chi_c$ (strongly nonlinear regime),
self-trapping effect arises and results in the formation of highly localized modes \cite{MolinaPRE2012,Molina1993-5SelfTrapping}. Moreover, Fig. 7
shows that this characteristic value $\chi_c$ increases with increasing SOC strength.

\begin{figure}[t!]
\includegraphics[angle=0,height=0.6\linewidth, width=0.9\linewidth]{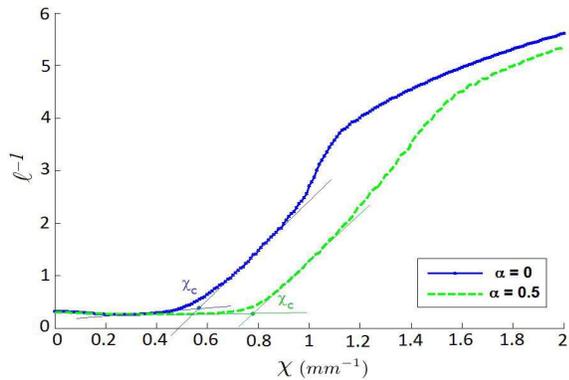}
\caption{(color online) Inverse of transverse localization length $\ell^{-1}$ versus nonlinear parameter $\chi$ for two values of relative coupling
strength $\alpha$. The excitation is at site $j_0=50$, and $\Delta=0.7$. }
\end{figure}

It is interesting to analyze the impact of the disorder strength $\Delta$ on the transverse localization in the case of
non-vanishing nonlinearity. This topic already attracted much attention in recent years \cite{Flach,Fishman}. In the absence of SOC, for $\chi \leq
0.6\ mm^{-1}$ (which is approximately equal to $\chi_c$; see Fig. 7 for $\alpha=0$) localization is stronger for increasing disorder level. This result is
consistent with our common sense, because the interference due to the disorder is an essential resource for Anderson localization. However, for
$\chi>\chi_c$, an increasing disorder level leads to less localization of the propagating wave packet. We attribute this anomalous result to
destructive interference effects due to the nonlinear Kerr phase shift. In the presence of SOC, for small values of $\alpha$ where localization is
not hindered, a similar behavior is observed. However, in this latter case, the anomalous result appears for larger value of $\chi$, which is due to
the growth of $\chi_c$ with increasing relative coupling strength $\alpha$. These findings are summarized in Fig. 8, where the effective width
$w_{eff}$ of the wave packet is plotted as a function of $z$, at relative SOC strengths $\alpha=0$ and $\alpha=0.5$, nonlinearity $\chi=0.7 \ mm^{-1}$, and
different disorder levels $\Delta=0.4, 0.5, 0.6, 0.7$. One can clearly see that for such high nonlinearities, the presence of SOC causes a decrease
of the effective width with increasing disorder level. If we increase in our simulations the nonlinearity to $\chi=1\ mm^{-1}$ -- which is larger than
$\chi_c$ for both $\alpha=0$ and $\alpha=0.5$ -- the anomalous result appears again. In this case, for both $\alpha=0$ and $\alpha=0.5$, high
disorder levels lead to less localized wave packets.

\begin{figure}[t!]
\includegraphics[angle=0,height=1.1\linewidth, width=0.9\linewidth]{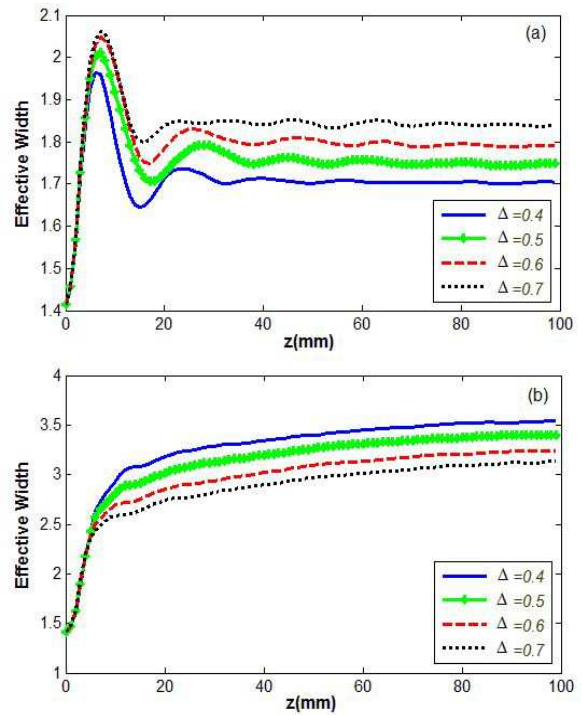}
\caption{(color online) The (dimensionless) effective width of wave packet as a function of $z$ for different disorder levels, with parameters $j_0=50$, $\chi=0.7\ mm^{-1}$
and (a) $\alpha=0$, (b) $\alpha=0.5$. }
\end{figure}

In order to complete our analysis, we repeat our analysis on the inverse TLL as a function $\alpha$ and $\chi$ for an excitation at
the edge of the lattice ($j_0 = 1$). The results are summarized in Fig. 9. We find all features of the bulk excitation (compare Fig. 2) also for the
edge excitation. The only difference between the cases of edge and bulk excitation is that the inverse TLL for the edge excitation is a highly
non-monotonic oscillating function of the relative SOC strength $\alpha$, for each nonlinear Kerr parameter $\chi$ (see Fig. 9). Though, the general trends prevails that localization gets weaker as $\alpha$ increases.

\begin{figure}[t!]
\includegraphics[angle=0,height=1\linewidth, width=0.9\linewidth]{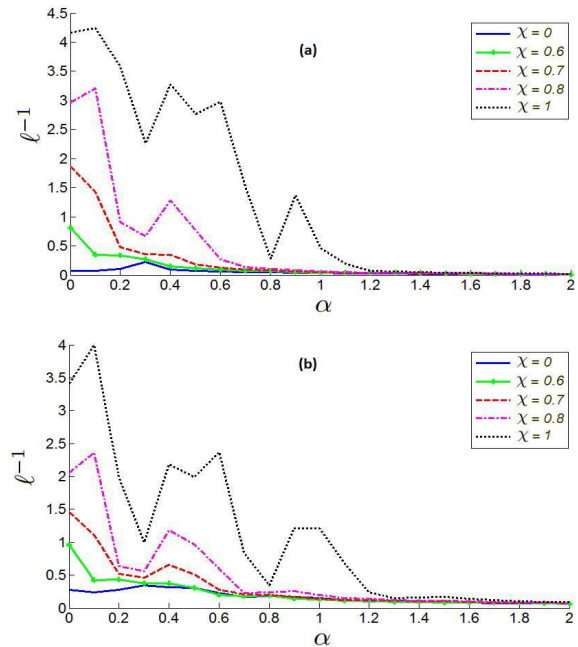}
\caption{(color online) Inverse of transverse localization length $\ell^{-1}$ versus relative coupling strength $\alpha$ for different values of the
nonlinear parameter $\chi$ ($mm^{-1}$).  The excitation is at site $j_0=1$. (a) $\Delta=0.4$, (b) $\Delta=0.7$.}
\end{figure}

In the linear regime ($\chi=0$) and in the absence of SOC ($\alpha=0$), the ratio of $\ell^{-1}$ for edge excitation ($j_0=1$) to
the value for bulk excitation ($j_0=50$) is approximately equal to $0.6$ and $0.9$, for disorder strengths $\Delta=0.4$ and $\Delta=0.7$,
respectively. This confirms that, in the linear regime, surface modes are less localized than bulk modes due to boundary repulsion
\cite{RepulsiveBoundary1,TL1DOffdiagonal}. This ratio increases by $50\%$ when the disorder level $\Delta$ grows from 0.4 to 0.7. Hence, for strong
disorder levels, the localization of surface and bulk modes is comparable \cite{MolinaPRE2012}. However, as can be seen in Figs. 2 and 9,
nonlinearity can reverse this effect. In this case, bulk modes become more extended than surface modes, which is in agreement with recent results
\cite{MolinaPRE2012}.

\section{Summary and Conclusion}

We have studied the effect of SOC on transverse localization in zigzag arrays of circular optical waveguides in the presence of Kerr nonlinearity.
Our simulations reveal that increasing next-nearest neighbor interaction hinders localization. We attribute this behavior to a
modified band structure of the Hamiltonian matrix in the presence of the second-order coupling. For bulk excitation the dependence of inverse TLL is
a monotonic function of the SOC, where for edge excitation the dependency is highly non-monotonic and oscillates for increasing SOC. Moreover, in the
absence of higher-order interactions, in the strongly nonlinear regime (i.e., when the nonlinear parameter is larger than some critical value),
counter-intuitively localization will decrease with increasing disorder level. The critical value, above which this anomalous behavior happens
increases with growing SOC.

\section*{Acknowledgment}

The authors would like to thank M. Khazaei Nezhad and Kh. Jafari for fruitful discussions.
This work was supported in part by
Sharif University of Technology's
Center of Excellence in Complex
Systems and Condensed Matter.
We also wish to thank the German Ministry of Education and Research (ZIK 03Z1HN31).

\end{document}